\documentclass[aps,twocolumn,twoside,superscriptaddress,longbibliography,nofootinbib, floatfix]{revtex4-1}

\usepackage{times}
\usepackage[utf8]{inputenc}
\usepackage{color}
\usepackage{amsmath}
\usepackage[pdftex,colorlinks=true,linkcolor=blue,citecolor=blue,urlcolor=blue]{hyperref}
\usepackage{qcircuit}
\usepackage{graphicx}

\newcommand{\ket}[1]{|#1\rangle}
\newcommand{\bra}[1]{\langle#1|}
\newcommand{\braket}[1]{\langle#1\rangle}

\newcommand{\poly}{\operatorname{poly}}

\newcommand{\be}{\begin{equation}}
\newcommand{\ee}{\end{equation}}

\begin{document}

\title{Compressed variational quantum eigensolver for the Fermi-Hubbard model}
\author{Ashley Montanaro}
\affiliation{PhaseCraft Ltd.}
\affiliation{School of Mathematics, University of Bristol, UK}
\author{Stasja Stanisic}
\email{stasja@phasecraft.io}
\affiliation{PhaseCraft Ltd.}

\begin{abstract}
    The Fermi-Hubbard model is a plausible target to be solved by a quantum computer using the variational quantum eigensolver algorithm. However, problem sizes beyond the reach of classical exact diagonalisation are also beyond the reach of current quantum computing hardware. Here we use a simple method which compresses the first nontrivial subcase of the Hubbard model -- with one spin-up and one spin-down fermion -- enabling larger instances to be addressed using current quantum computing hardware. We implement this method on a superconducting quantum hardware platform for the case of the $2 \times 1$ Hubbard model, including error-mitigation techniques, and show that the ground state is found with relatively high accuracy.
\end{abstract}

\date{\today}

\maketitle

\setlength{\parskip}{3pt}

The Fermi-Hubbard model is one of the cornerstones of condensed-matter physics and a fundamental tool for the study of strongly correlated electron systems~\cite{hubbard63,hubbard13}. However, solving the model is a very significant challenge, both theoretically and numerically~\cite{yamada05,leblanc15}. This has motivated the suggestion that quantum computers may be able to address instances of the model beyond the capacity of classical methods.

Here our focus is on solving the Hubbard model in the sense of finding the ground state of the fermionic Hamiltonian
\be \label{eq:hubbard} H = -t \sum_{\langle i, j\rangle,\sigma} (a_{i\sigma}^\dag a_{j\sigma} + a_{j\sigma}^\dag a_{i\sigma}) + U \sum_k n_{k\uparrow}n_{k\downarrow}, \ee
where the notation $\langle i, j \rangle$ denotes sites that are adjacent on a lattice, and $\sigma \in \{\uparrow,\downarrow\}$. It is plausible that this problem could be solved using the variational quantum eigensolver~\cite{peruzzo14,mcclean16} (VQE) approach, a method based on the use of a classical algorithm to optimize over the space of quantum circuits for producing the ground state. VQE has been previously studied in the context of the Hubbard model, with promising results~\cite{wecker15,reiner19,Verdon2019,Dallaire-Demers2018,cai19,cade20}.

To solve a fermionic system on a quantum computer, an encoding method must be used to represent the system in terms of qubits. Usually, in variational methods the Hubbard Hamiltonian (\ref{eq:hubbard}) is expressed in second-quantised form, and then transformed via the Jordan-Wigner transform, or another method~\cite{bravyi02,ball05,verstraete05,derby20}, which enforces the fermionic antisymmetry. These methods represent a system with $N$ fermionic modes using $N$ qubits, or more. For a Hubbard model system with $n$ sites, this corresponds to the need for at least $2n$ qubits. A detailed analysis of the complexity of VQE applied to the Hubbard model was developed in~\cite{cade20}, which concluded that instances beyond the capacity of classical exact diagonalisation might be accessible using quantum circuits on 50 qubits and with depth less than 1000. While a significant reduction compared with previous estimates, this is still beyond the reach of today's quantum computers.

The Hubbard Hamiltonian preserves fermionic occupation number and spin type, implying that its ground state can be found by restricting to a subspace with particular occupation number and spin type. Here we take advantage of this feature to handle a particularly simple case exponentially more efficiently: the case where we have one spin-up electron and one spin-down electron. This is the first non-trivial case of the Hubbard model, in the sense that if there are fewer than one spin-up or spin-down electrons, the onsite term vanishes and we are left with a system of noninteracting fermions.

Rather than seeking to find efficient quantum circuits that generate states of $2n$ qubits within a subspace of a fixed occupation number within time $\poly(n)$~\cite{gard19}, here we compress this space down to only $\poly\log(n)$ qubits, and use efficient quantum circuits of size $\poly\log(n)$ to operate within this space. The representation we use is similar to first quantisation, but with some differences which we outline below. It can be seen as a simple variant -- specialised to the case of the Hubbard model -- of the configuration interaction (CI) matrix method from quantum chemistry, which was introduced in the context of quantum algorithms in~\cite{toloui13,babbush18a}. Other efficient representations of fermionic systems with occupation number constraints, which are more general and complex than the representation here, have been developed previously~\cite{moll16,bravyi17,babbush19}.

We used the VQE algorithm to optimize over circuits operating on this space, using the Hamiltonian variational ansatz~\cite{wecker15} within the optimized framework developed in~\cite{cade20}. We implemented the algorithm for the simplest nontrivial case of the Hubbard model -- a $2\times 1$ lattice -- using commercially-available cloud quantum computing hardware (the Rigetti Aspen-4 and Aspen-7). We compared the performance of solving a compressed instance on 2 qubits with solving an uncompressed instance on 4 qubits, taking into account the effect of error-detection~\cite{cade20} and readout noise mitigation~\cite{kandala17,endo18,maciejewski19} procedures. The VQE algorithm was not able to find an approximate ground state of the uncompressed instance, but was able to find a relatively high-accuracy approximation to the ground state of the compressed instance.

We then used this approximate ground state to compute a physically meaningful quantity: the double occupancy $\braket{\frac{1}{n}\sum_k n_{k\uparrow}n_{k\downarrow}}$, which provides information about the Mott-insulating character of the state~\cite{parcollet04,jordens08}. We computed the ground-state energy and the double occupancy for a varying $U$ parameter between 0.1 and 4. The median absolute error in energy is $\approx 6.5 \times 10^{-2}$, while the median absolute error in double occupancy is $\approx 5.7 \times 10^{-3}$. 
Notably, this is substantially lower than the 2-qubit gate infidelity of the quantum processor ($\approx 4\%$).

In the special case of a $2\times 1$ lattice, the ground state within the subspace of occupation number $1+1$ is actually the ground state of the full Hamiltonian $H$ in (\ref{eq:hubbard}). These results thus demonstrate the solution of an instance of the Hubbard model using VQE on quantum computing hardware.

We remark that, for this $2\times 1$ case, the encoding we use turns out to be the same as one introduced in~\cite{moll16} and explicitly calculated for the $2\times 1$ Fermi-Hubbard model; the same encoding was used to find the ground state of the $2\times 1$ Fermi-Hubbard model via a discretised adiabatic algorithm on an ion-trap quantum computer~\cite{linke18}. Recent work has addressed the related problem of solving the Hubbard model on the Bethe lattice in infinite dimensions, in the context of dynamical mean-field theory, by finding ground states and simulating time-evolution of impurity models using quantum computing hardware with 4 qubits~\cite{keen20,rungger19}.

Different notions of compressed quantum simulation have been previously studied. A particularly fruitful notion exploits a close relationship between noninteracting fermionic Hamiltonians of the form $\sum_{i \neq j} c_{ij} a_i^\dag a_j$, matchgates, and log-space quantum computation~\cite{terhal02,jozsa10}. This allows the time-evolution of certain systems on $n$ qubits (such as the 1d Ising model with transverse field and the XY model) to be simulated using a system of $O(\log n)$ qubits~\cite{kraus11,boyajian13}. Using this idea, Ising systems with transverse field have been simulated using 2 qubits on a cloud-based superconducting quantum processor~\cite{hebenstreit17} and 5 qubits on an NMR quantum simulator~\cite{li14}.

Another notion of compression in quantum simulation is where the Hamiltonian in question can be diagonalised by an efficient quantum circuit. Taking advantage of this capability, ground state and time dynamics simulation of the antiferromagnetic Ising model with transverse field have been studied on quantum computers with 4 qubits~\cite{cerveralierta18}, showing some qualitative agreement with theoretical results, but significant quantitative differences due to hardware limitations.

Finally, one can consider \emph{automatic} compression of quantum systems, via the concept of quantum autoencoders, which are a method to learn compressed representations of quantum states~\cite{romero17a}. Quantum autoencoders have been applied to the Hubbard model~\cite{romero17a}, compressing (for example) the ground state of a $2\times 1$ instance to 2 qubits or even 1, with low error (albeit inexactly).

\section{Compressed representation}
\label{sec:compress}

Our starting point is the well-known Jordan-Wigner transformation. In this transformation, each fermionic mode corresponds to a qubit. Each hopping term $h_{ij}$ between a pair of modes $i$ and $j$ ($i < j$) maps to a qubit operator via
\[ a_i^\dag a_j + a_j^\dag a_i \mapsto \frac{1}{2}(X_iX_j + Y_iY_j) Z_{i+1} \cdots Z_{j-1}. \]
Each onsite term acting on modes $i$ and $j$ maps to a qubit operator via 
\[ a_i^\dag a_i a_j^\dag a_j \mapsto \frac{1}{4}(I - Z_i)(I - Z_j), \]
whether or not qubits $i$ and $j$ are adjacent in the Jordan-Wigner encoding.

Assume we have $n$ sites with some interaction graph $G=(V,E)$ where $|V|=n$, $|E|=m$ (for example, a rectangular lattice), corresponding to $2n$ fermionic modes. We choose to order the fermionic modes such that all the spin-up modes come before all the spin-down modes. Then we define a basis for the modes of each spin, of the form $\{\ket{e_i}:i \in \{1,\dots,n\}\}$, where $e_i \in \{0,1\}^n$ is the bit-string of Hamming weight 1 which has a 1 at the $i$'th position. The space of states of occupation number 1 for each spin-type is then spanned by the basis $\{\ket{e_i}\ket{e_j}:i,j \in \{1,\dots,n\}\}$. Writing
\[ H = -t (H_{hop} \otimes I + I \otimes H_{hop}) + U H_{os}, \]
we have, for each hopping term $h_{ij}$,
\[ \braket{e_k|h_{ij}|e_k} = \begin{cases} 1 & k < i \text{ or } k>j\\ -1 & i < k < j\\
0 & k = i \text{ or } k=j
\end{cases}, \]
and for $k \neq l$,
\[ \braket{e_k|h_{ij}|e_l} = \begin{cases} 1 & k=i, l=j \text{ or } k=j, l=i \\
0 & \text{otherwise.} \end{cases} \]
Summing over $i$ and $j$ such that $(i,j) \in E$, we get
\begin{multline*}
\braket{e_k|H_{hop}|e_k} = |\{(i,j) \in E: k \not\in\{i,j\} \}|\\
- 2|\{(i,j) \in E: i < k < j \}|,
\end{multline*}
and
\[ \braket{e_k|H_{hop}|e_l} = \begin{cases} 1 & (k,l) \in E\\ 0 & \text{otherwise.} \end{cases} \]
The onsite term can be calculated directly as
\[ \bra{e_i}\braket{e_j|H_{os}|e_k}\ket{e_l} = \begin{cases} 1 & i=j=k=l \\ 0 & \text{otherwise.} \end{cases} \]
For each Hamiltonian $H$, $H_{hop}$, $H_{os}$, we use a superscript $C$ to denote the corresponding ``compressed'' Hamiltonian projected onto the occupation number 1 subspace (for each spin type).
Note that the off-diagonal entries of $H^C_{hop}$ are given by the adjacency matrix of $G$, but that in general the diagonal entries depend on the ordering we chose in the Jordan-Wigner transform. In the case where $G$ is a line, so $E=\{(1,2),(2,3),\dots,(n-1,n)\}$, and we choose the natural Jordan-Wigner ordering, we see that $\braket{e_k|H^C_{hop}|e_k} = n-2$ for all $k$.

We can associate each $n$-qubit state $\ket{e_i}$ with a  state $\ket{i}$ of $p := \lceil \log_2 n \rceil$ qubits, corresponding to writing $i$ in binary. This then gives us an exponentially compressed representation, with respect to the original $2n$ qubits. States of the $2\lceil \log_2 n\rceil$ qubits of the compressed system are of the form $\sum_{i,j=1}^n \alpha_{ij} \ket{i}\ket{j}$. Note that any such state corresponds to a valid physical state. The steps required for the VQE algorithm can be implemented in time $\poly \log(n)$, rather than $\poly(n)$, for arbitrary sparse interaction graphs, such as lattices (see Section~\ref{sec:general} for a discussion).

We remark that for certain interaction graphs (such as a line), it would be possible to further reduce the number of qubits used by taking advantage of additional symmetries of the graph. However, this would lead to a more complicated representation and would also not allow for nonuniform local terms.

\textbf{Related representations.} This representation is similar to first quantisation, but there are some differences. In first quantisation, a state of two fermions in a system of $N$ fermionic modes can be written as
\be \label{eq:firstquantisation} \sum_{i < j} \alpha_{ij} \frac{1}{\sqrt{2}}(\ket{ij} - \ket{ji}), \ee
where $\sum_{i < j} |\alpha_{ij}|^2 = 1$. In the case of the Hubbard model on a lattice of $n$ sites, this would correspond to a quantum state of $2\lceil \log_2 (2n) \rceil = 2\lceil \log_2 n \rceil + 2$ qubits, which is less efficient. In addition, preparing a state of the form of (\ref{eq:firstquantisation}) is more complex than the states of the form $\sum_{i,j} \alpha_{ij} \ket{i}\ket{j}$ that we consider. It is interesting to note that, in our setting, fermionic antisymmetry is handled via the projected Hamiltonian $H^C$, rather than being a property of the state space.

The representation we use is closely related to one based on the configuration interaction (CI) matrix representation from quantum chemistry, which was introduced in the context of quantum computing in~\cite{toloui13,babbush18a}. These representations are also based on restricting a second-quantised fermionic Hamiltonian to a particular occupation number subspace. The space of $N$ modes with total occupation number $\eta$ has dimension $\binom{N}{\eta}$. Basis states for this subspace can be encoded as qubits either as a tensor product of $\eta$ $\lceil \log_2 N \rceil$-qubit registers (similarly to the method we use, and to first quantisation) or as a $\lceil \log_2 \binom{N}{\eta} \rceil$-qubit register directly. Here we save a qubit or two compared with the more general representations described in~\cite{toloui13,babbush18a} by using that the Hubbard Hamiltonian preserves spin-type.

\textbf{The $2 \times 1$ case.} We will focus on the smallest nontrivial case of the Hubbard model: a $2\times 1$ lattice. With respect to the basis $\{\ket{0},\ket{1}\}$,
\[ H^C_{hop} = \begin{pmatrix} 0 & 1 \\ 1 & 0 \end{pmatrix} = X ,\;\; H^C_{os} = \begin{pmatrix} 1 & 0 & 0 & 0\\ 0 & 0 & 0 & 0\\ 0 & 0 & 0 & 0\\ 0 & 0 & 0 & 1 \end{pmatrix} = \frac{1}{2}(I + Z \otimes Z). \]
Therefore, the overall Hamiltonian $H^C$ is of the form
\begin{eqnarray*} \label{eq:hc} H^C &=& -t(X \otimes I + I \otimes X) + \frac{U}{2}(I + Z \otimes Z)\\
&=& \begin{pmatrix} U & -t & -t & 0\\ -t & 0 & 0 & -t\\ -t & 0 & 0 & -t\\ 0 & -t & -t & U \end{pmatrix}
\end{eqnarray*}
when restricted to the subspace with one spin up, one spin down (the same representation was derived by a different approach in~\cite{moll16}). The ground state of $H^C_{hop}$ is straightforward to prepare as $\frac{1}{\sqrt{2}}(\ket{0} - \ket{1})$.

$H^C$ is simple enough to be diagonalised analytically. The ground state is
\[ \frac{\alpha}{\mathcal{N} \sqrt{2}}(\ket{00}+\ket{11}) + \frac{\beta}{ \mathcal{N} \sqrt{2}}(\ket{01}+\ket{10}) \label{eq:groundstate} \]
with $\alpha = 4$, $\beta = U + \sqrt{U^2 + 16}$, $\mathcal{N} = \sqrt{\alpha^2 + \beta^2}$, corresponding to energy $E = U/2 - \sqrt{U^2/4 + 4t^2}$.

\section{Variational quantum eigensolver}
\label{sec:vqe}

Our goal is to find the ground state of $H$ using the VQE framework. This approach uses a classical optimizer to optimize over a family (``ansatz'') of quantum circuits. The aim is to find a circuit that produces a state with minimal energy with respect to $H$, where the energy is estimated using a quantum computer. There are many variants of VQE; here we used an approach analysed in~\cite{cade20} and found to be effective.

The family of circuits used is the Hamiltonian variational ansatz presented in~\cite{wecker15}. The Hubbard Hamiltonian $H$ in (\ref{eq:hubbard}) is split into horizontal, vertical and onsite parts, each of whose terms pairwise commute. The circuit begins by preparing the ground state of the quadratic part of $H$ (equivalently, taking $U=0$), which can be done efficiently by diagonalising the matrix specifying the quadratic part of $H$ using Givens rotations~\cite{jiang2018quantum}. Then the circuit consists of a number of layers, each of which includes time-evolution according to each of the parts of the Hamiltonian in turn. The lengths of time each part evolves for are the parameters to be optimized classically. This split into parts also provides a natural approach to measuring the energy of the trial state, by combining estimates of the energy of each part.

Importantly, because all the operations used in this ansatz correspond to time-evolution according to terms of the Hubbard Hamiltonian, they preserve occupation number and spin-type. This means that this ansatz can immediately be applied in our compressed context.

\begin{figure*}[t]
    \centering
    \[
        \Qcircuit @C=1em @R=1em @!R {
            & \gate{X} &  \ctrl{1} & \gate{RY(-\frac{\pi}{2})} & \ctrl{1} & \push{\rule{0em}{2em}} \qw &  \push{\rule{0em}{2em}} \qw &  \gate{RZ(\phi)} & \qw  & \ctrl{1} & \gate{RX(\theta)} & \ctrl{1}   &  \qw & \ctrl{1} & \gate{H} & \ctrl{1}   &\push{\rule{0em}{2em}}\qw\\
            & \qw & \targ & \ctrl{-1} & \targ & \qw & \qw & \qw &    \gate{RZ(\phi)}  &  \targ & \ctrl{-1} & \targ & \qw  &  \targ & \ctrl{-1} & \targ &\qw\\
            & \gate{X} & \ctrl{1} & \gate{RY(-\frac{\pi}{2})} & \ctrl{1} & \qw & \qw & \ctrl{-2}  & \qw &   \ctrl{1} & \gate{RX(\theta)} & \ctrl{1}   & \qw &  \ctrl{1} & \gate{H} & \ctrl{1}  &\qw\\
            &  \qw & \targ & \ctrl{-1} &\targ & \push{\rule{0em}{1em}} \qw & \push{\rule{0em}{1em}} \qw  &\qw  & \ctrl{-2} &  \targ & \ctrl{-1} & \targ  & \qw & \targ & \ctrl{-1} & \targ  & \push{\rule{0em}{1em}}\qw \gategroup{1}{1}{4}{6}{.7em}{--} \gategroup{1}{7}{4}{13}{.7em}{--} \gategroup{1}{14}{4}{17}{.7em}{--}
        }
    \]
    \caption{Uncompressed circuit with initial state preparation, ansatz, and measurement highlighted (where the last step only happens for hopping term measurements)}
    \label{fig:circuit_full}
\end{figure*}
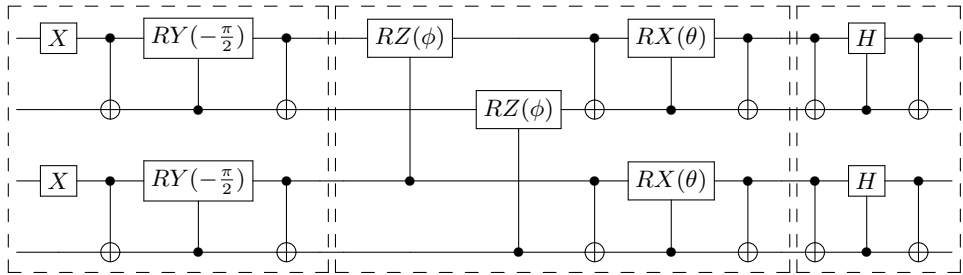

It was found in~\cite{cade20} that a single ansatz layer is sufficient to find the ground state of the $2 \times 1$ Hubbard model. As the $2\times 1$ case only has one horizontal term and one onsite term, the variational ansatz has two parameters. This single layer consists of onsite gates between the appropriate pairs, followed by the horizontal hopping gate between the appropriate pairs (see Figure~\ref{fig:circuit_full}).
This single layer then results in a state of the form
\begin{equation*}
    e^{i \theta H_\mathrm{hop}} e^{i \phi H_\mathrm{os}} \ket{\Psi_{\mathrm{ini}}}
\end{equation*}
where $H_\mathrm{hop}$ is the hopping terms, $H_\mathrm{os}$ is the onsite terms, $\ket{\Psi_{\mathrm{ini}}}$ is the initial state, and $\theta$ and $\phi$ are the parameters we are optimizing over.
Finally, there are only two types of measurements that need to be carried out: the onsite measurement and the horizontal pairs measurement.
To carry out the horizontal pairs measurement we need a measurement preparation step which transforms into the $\frac{1}{2}(XX+YY)$ basis.

The algorithm used to carry out the optimization step of the VQE is SPSA~\cite{spsa}, as in~\cite{kandala17,cade20} it was found to be capable of coping with the type of noise we expect on a non-error corrected quantum processor.
We use hyperparameter choices as described in~\cite{cade20}, and we implement both the standard SPSA as well as an enhancement used in that work where initially coarse function evaluations are used involving fewer energy measurements, before more precise function evaluations.

The full sequence of the algorithm is then:
\begin{itemize}
\item Assign an initial guess for the angles to be used in the ansatz.
\item Find the necessary energy evaluation on the quantum processor for the given parameters (running the circuit for onsite and horizontal measurement a certain number of times to get the energy measurement).
\item Adjust the parameters based on the energy evaluation.
\item Iterate until stopping number of iterations.
\item Take a single detailed data point.
\end{itemize}
At the end of the algorithm we expect to have the ground energy and the ground state, allowing us to also calculate some more physical properties such as probability of double occupancy.

\section{Implementation and experimental results}

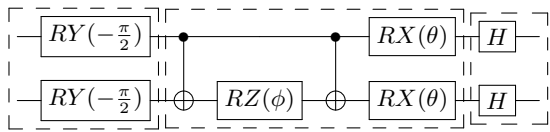
\begin{figure*}[t]
    \centering
    \[
        \Qcircuit @C=1em @R=1em @!R {
            & \gate{RY(-\frac{\pi}{2})} & \ctrl{1} & \qw &   \ctrl{1} & \gate{RX(\theta)} &  \gate{H} &\qw\\
            & \gate{RY(-\frac{\pi}{2})} & \targ & \gate{RZ(\phi)} & \targ  & \gate{RX(\theta)} &  \gate{H} &\qw \gategroup{1}{1}{2}{2}{.7em}{--} \gategroup{1}{3}{2}{6}{.7em}{--} \gategroup{1}{7}{2}{8}{.7em}{--}
        }
    \]
    \caption{Compressed circuit, with same highlights as uncompressed circuit}
    \label{fig:circuit_two_qubits}
\end{figure*}

We employ two noise reduction strategies.
The first one is error detection as described in~\cite{cade20}, where the total number of fermions detected in the outcome has to be preserved when compared with the initial state (so there should be two fermions present). If an incorrect number of fermions is measured, then that run is discarded.
We go a step further here, and check that not only the total number of fermions is preserved, but also the number of fermions in each spin subspace.
This correction is only carried out for the uncompressed implementation, as the compressed implementation already encodes the states in a way that preserves these quantities.

The other is a common error correction procedure to handle readout errors~\cite{kandala17,endo18,maciejewski19}.
Before running the algorithm, we sample the noise by producing each computational basis state, then measuring in the computational basis.
In an ideal situation, the only outcome should be the bitstring corresponding to that computational basis state.
Let the ideal distribution be $p$, and let the measured distribution be $\tilde{p}$.
We assume that this noise map $\mathcal{N}$ acts as $\mathcal{N} p = \tilde{p}$.
We can then estimate this $\mathcal{N}$, invert it and apply it to the measured distribution to get the ideal distribution.

As mentioned in Section~\ref{sec:vqe}, we implement both a standard and three-stage SPSA algorithm.
For the standard SPSA algorithm, we use a fixed number of iterations (based on visual inspection of ``flattening'' of the energy value) as well as a fixed number of energy measurements (10,000).
We keep the number of standard SPSA iterations fixed to 175 (see Figure~\ref{fig:vqe}).
The modified SPSA algorithm is carried out in three stages: coarse measurement using 250 iterations of 100 energy measurements, intermediate using 50 iterations of 1,000 energy measurements, and fine using 25 iterations of 10,000 measurements.
The circuit used for energy estimates is run using the active reset feature of the Rigetti processors (which allows for quicker but possibly less precise measurements).
After the final iteration of both the standard and the three-stage SPSA, the resulting parameters are used for a more precise measurement using 10,000 energy measurements, switching off the active reset feature available on Rigetti processors, and with a noise inversion matrix calculated immediately before the measurements (unlike for the rest of the SPSA where the noise inversion matrix is calculated once at the beginning).
The starting parameters are $(\phi, \theta) = (1, 1)$.

\subsection{The original circuit using four qubits}
\label{sec:original_circuit}

We implemented both the compressed and uncompressed approaches on the Rigetti Aspen-7-4Q-D processor. This device has exactly four sites in a square configuration, which enables all the required gates to be implemented across neighbouring qubits with no need for swapping.

Figure \ref{fig:circuit_full} illustrates the uncompressed VQE circuit on 4 qubits. 
The direct implementation of gates from Figure~\ref{fig:circuit_full} using just the Quil compiler optimization is $\sim$43 with multiqubit depth (in this case, CZ gates) of 8 for the onsite measurement circuit, and $\sim$46 with multiqubit depth of 8 for the horizontal measurement circuit (there is a slight variation from compilation to compilation due to compiler optimization, but this is only by a gate or two).

To improve the initial depth of the circuit based on the gates as we have defined them in Figure~\ref{fig:circuit_full}, we do some manual optimization of the gates in the native gateset.
These manually optimized gates in combination with Quil compiler optimization give total gate depth of 30 with 6 multiqubit depth for onsite measurements, and 35 with 6 multiqubit depth for horizontal measurements (with possible further optimization available here, by merging gates needed for basis change with the horizontal hopping gates at the end of the circuit as both of these sets of gates are applied to the same qubits -- see Figure~\ref{fig:circuit_full})

\begin{figure*}[t]
    \centering
    \includegraphics[width=0.99\linewidth]{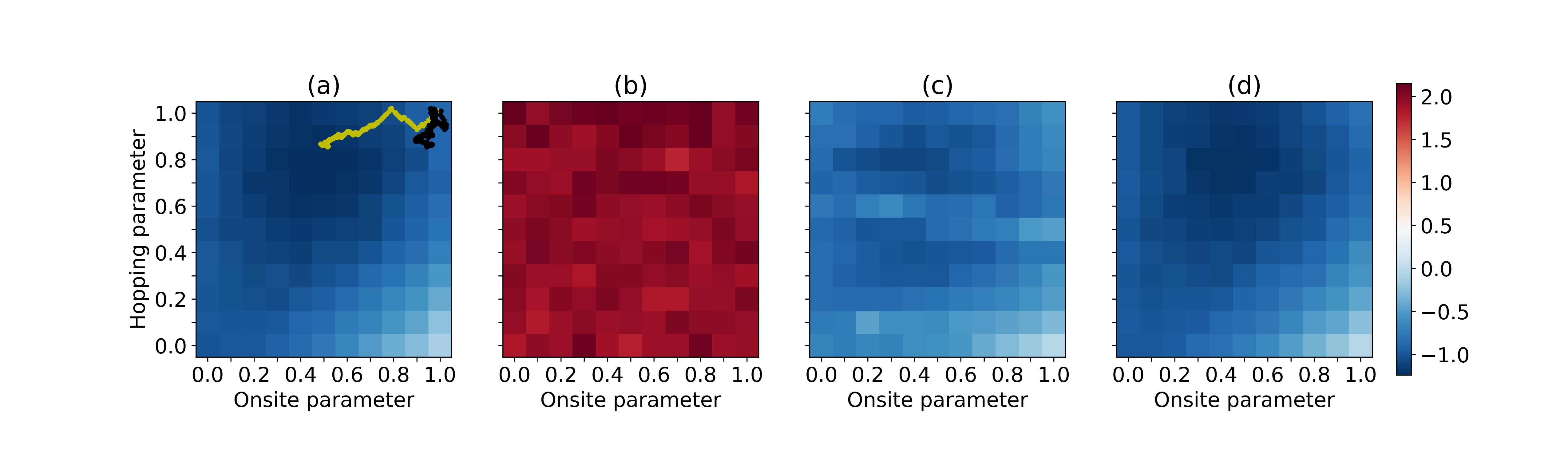}
    \caption{Heatmaps of corrected energy values for the a) simulated, b) uncompressed implementation, and (c and d) compressed implementation. The settings in Eq.~\ref{eq:hubbard} are taken to be $U=2$ and $t=1$. The sweep is over the ansatz onsite ($\phi$) and horizontal ($\theta$) parameters from $[0, 1]$ in $0.1$ increments. The actual energy of the ground state is $-1.23607$ with ansatz parameters of $(0.46365, \pi/4)$ as we can notice from the simulated heatmap (a). a) Simulated heatmap. The superposed yellow data points show the walk over the parameters of one of the runs of VQE using Aspen-7-4Q-D chip and the compressed implementation (see also Figure~\ref{fig:vqe}). The black data points show the same walk but for uncompressed implementation. b) Heatmap of uncompressed circuit on Aspen-7-4Q-D with no distinct minimum (values after error correction). Note that the energy values in this heatmap are positive even after correction -- compare this with executing just the identity circuit after the initial state preparation, which should give an energy value of $-1.0$. c) Heatmap of the compressed circuit on qubits 21 and 36 of Aspen-7-4Q-D showing some similarities with the simulated heatmap in a). d) Heatmap of the compressed circuit on Aspen-4-2Q-C showing a similar landscape to that in a).}
    \label{fig:heatmaps}
\end{figure*}

The first experiment we carried out is a comparison of the energy landscape for simulated and chip data using a parameter sweep over the onsite and horizontal gate parameters, for the case of $U=2$, $t=1$.
The final results, after error detection and correction are given in Figure~\ref{fig:heatmaps}.
Comparing the two heatmaps, we can see immediately that the simulated heatmap has a distinct minimum in a concave part of the landscape, while the heatmap from the QPU looks much noisier, with a substantially different landscape.

The second experiment we carried out is an attempt at VQE as described in Section~\ref{sec:vqe}.
In Figure~\ref{fig:vqe}, we show the results based on 3 runs of the VQE with and without noise correction.
As expected from the heatmap with no definite minmum, the VQE in Fgure~\ref{fig:vqe} demonstrates lack of change in energy value over iterations (beyond noise).
Moreover, the noise correction does not seem to help, or even worse, makes the energy values higher.
This is likely due to the fact that there is no true data in this noise, demonstrating further work is needed to make the algorithm viable.

\subsection{The compressed circuit using two qubits}
We then implemented the compressed version of the algorithm as outlined in Section~\ref{sec:compress} and circuit as shown in Figure~\ref{fig:circuit_two_qubits}.
To compare the performance of the four (uncompressed) and two (compressed) qubit instance, we run the same tests on the same chip as in the previous section (Aspen-7), but out of the four qubits used there we pick the two best performing ones. The best performing set of qubits are picked by checking their gate fidelity as well as performance on some initial VQE tests.
There is no noise detection based on the number of fermions detected, as this information is now encoded into our state.
However, we can still implement the standard readout correction.

\begin{figure*}[t]
    \centering
    \begin{minipage}[t]{0.49\linewidth}
        \includegraphics[width=1.0\linewidth]{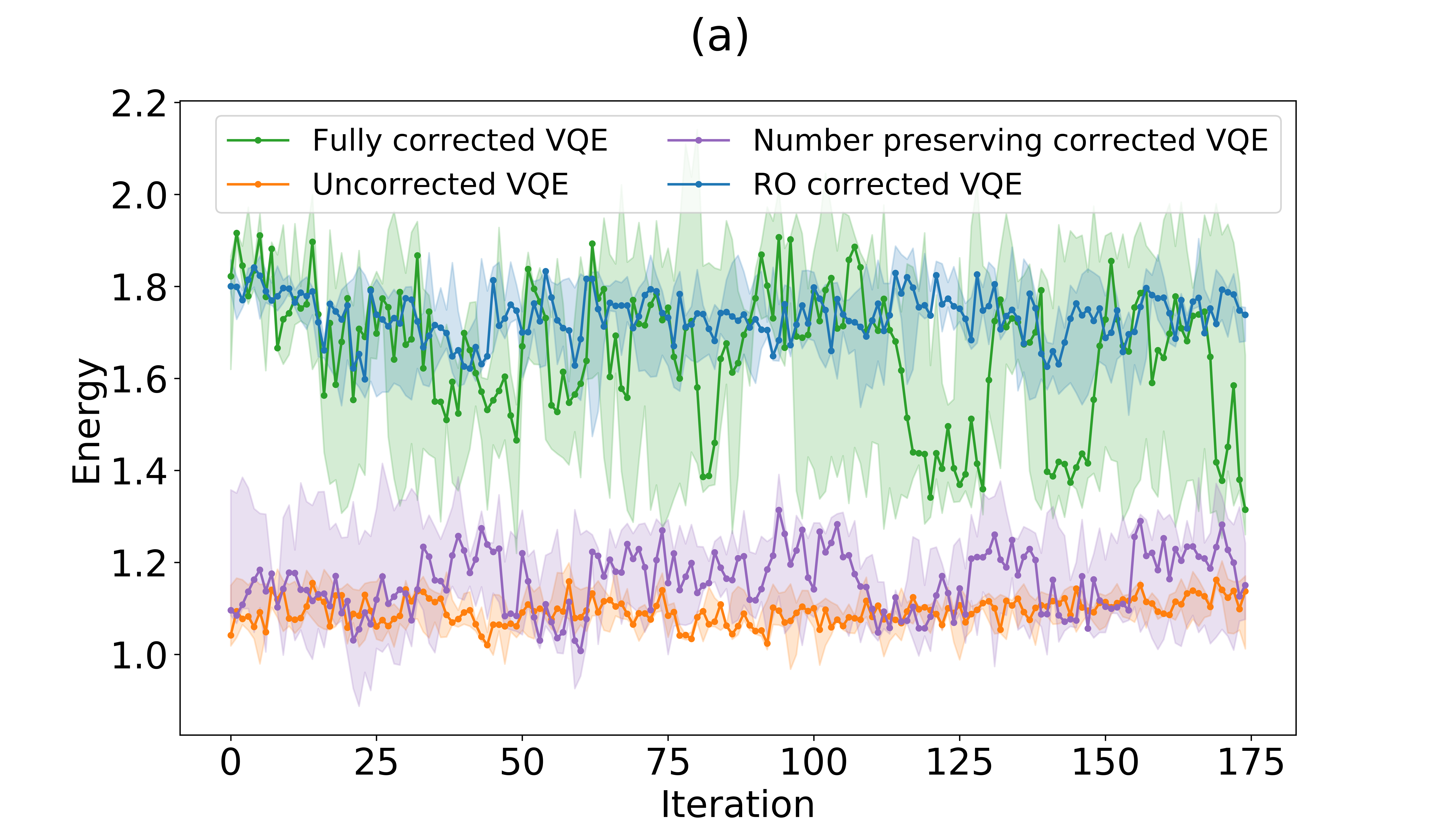}
    \end{minipage}
    \begin{minipage}[t]{0.49\linewidth}
        \includegraphics[width=1.0\linewidth]{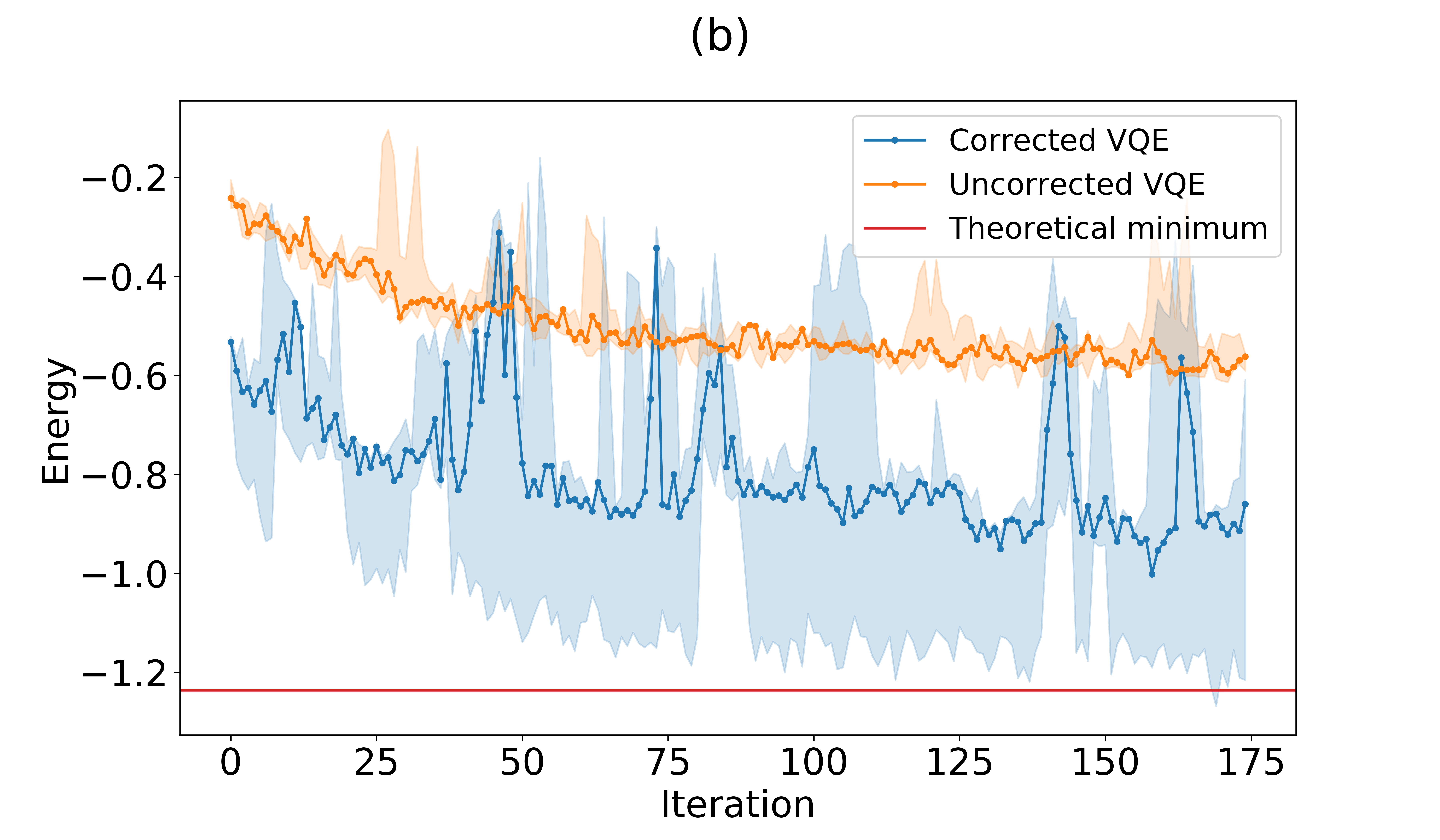}
    \end{minipage}\\
    \begin{minipage}[t]{0.49\linewidth}
        \includegraphics[width=1.0\linewidth]{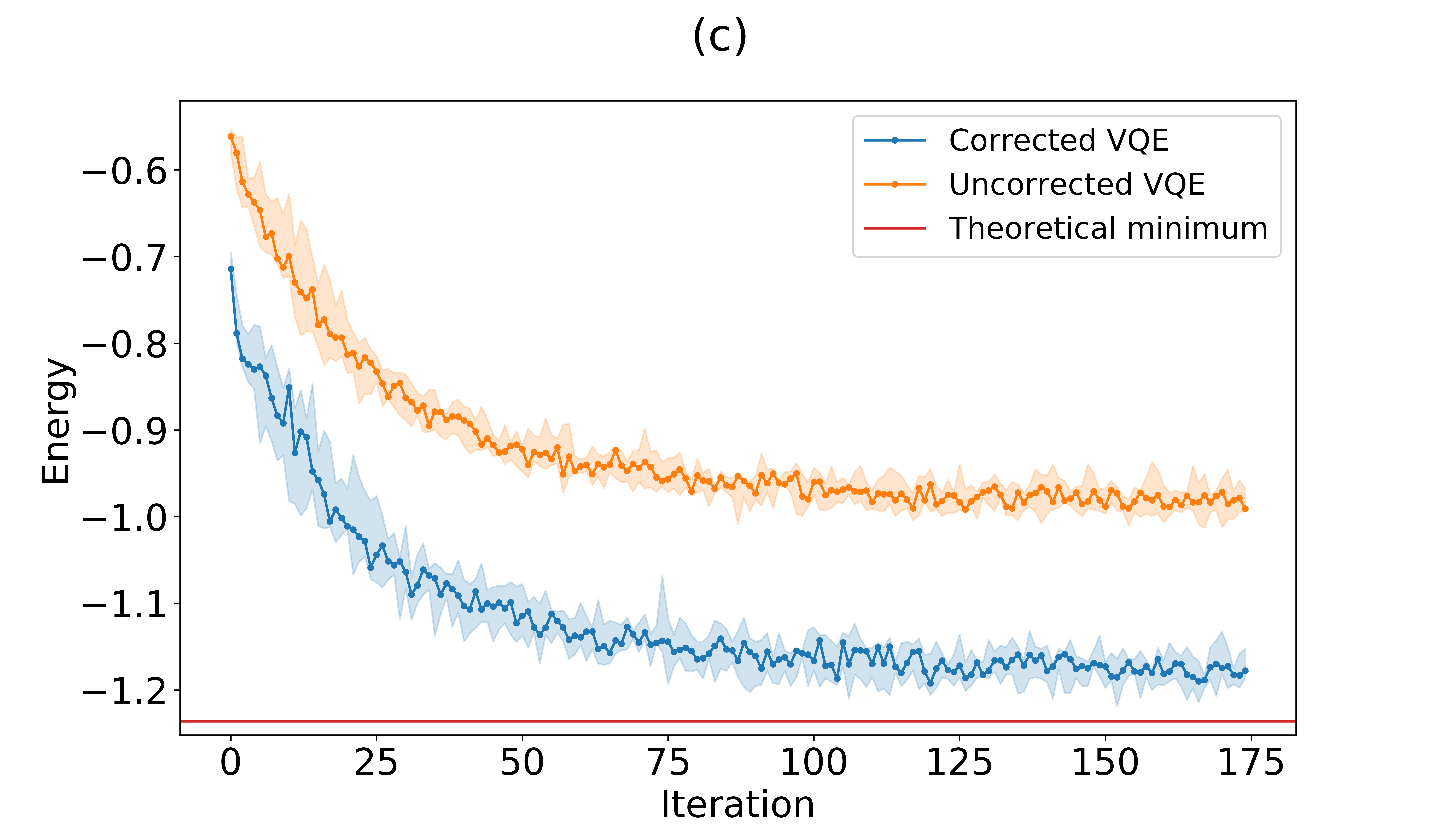}
    \end{minipage}
    \begin{minipage}[t]{0.49\linewidth}
        \includegraphics[width=1.0\linewidth]{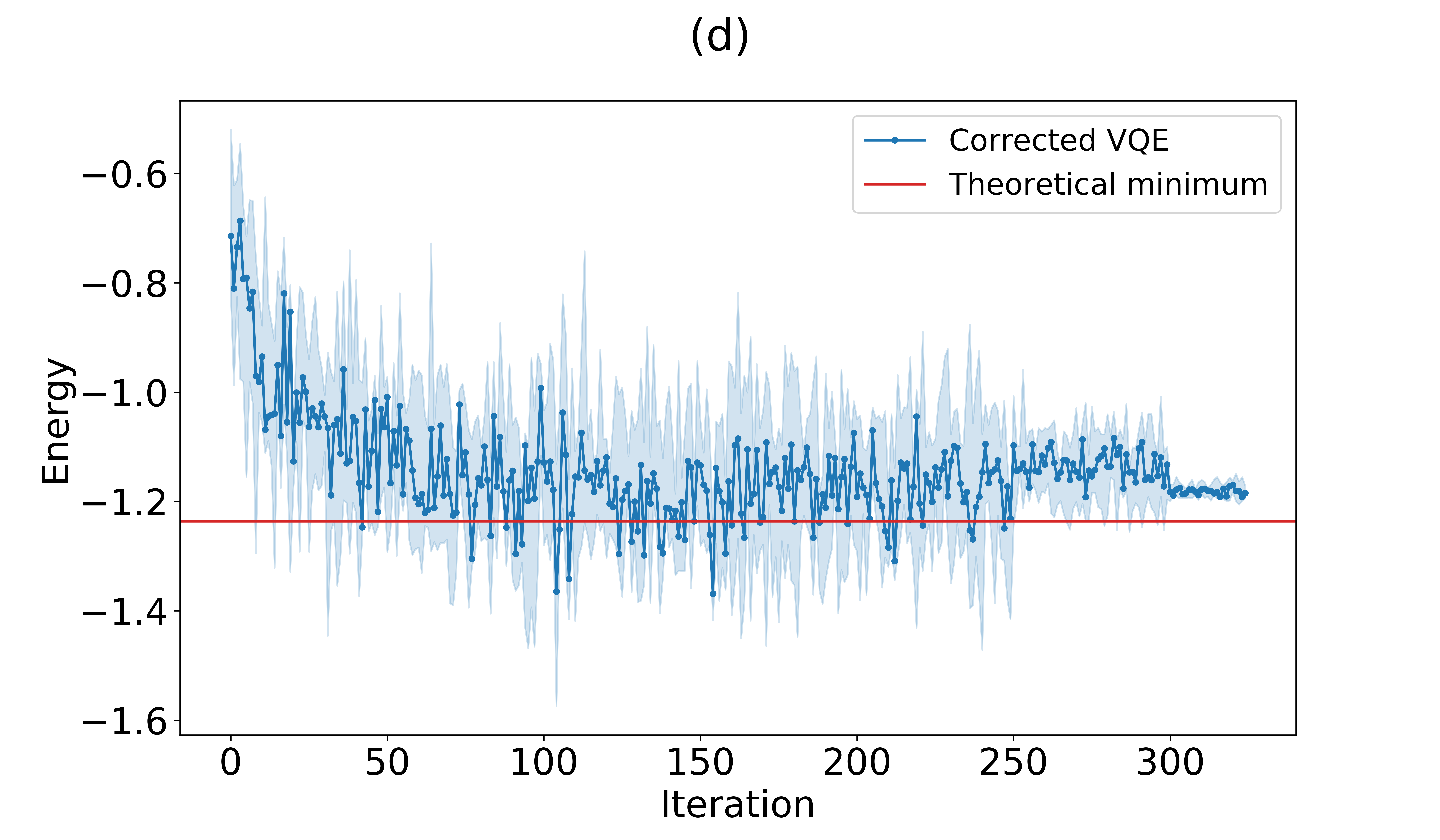}
    \end{minipage}
    \caption{Results of VQE performed on processors Aspen-7-4Q-D and Aspen-4-2Q-C with uncompressed (a) and compressed (b, c, d) implementations, using original (a, b, c) and three-stage (d) SPSA algorithm. The graphs show energy values at each iteration of the SPSA algorithm. The orange data points show uncorrected VQE, the blue datapoints show readout corrected VQE. The theoretical minimum (red) is the predicted ground energy for $U=2$ and $t=1$ of $E_{\mathrm{min}} = -1.23607$. Data shown gives a median of three (five) runs and min/max envelope on Aspen-7 (Aspen-4). a) The results of uncompressed implementation on Aspen-7-4Q-D. The green datapoints show VQE executed using both readout and occupation number preserving correction. The purple datapoints show VQE executed using occupation number preserving correction only. It is peculiar how the error correction actually makes the energy values worse in certain cases -- this is possibly another indicator that the data is too noisy and the signal too weak to be corrected. The median final fully corrected value is $1.40$.
    b) VQE executed using the compressed circuit on Aspen-7-4Q-D, qubits 21 and 36. The median final corrected value is $-1.06$. c) VQE executed using the compressed circuit on Aspen-4-2Q-C. The median final corrected value is $-1.17$. d) The results of the three-stage SPSA algorithm: 250 iterations of 100 energy measurements, 50 iterations of 1,000 energy measurements, and 25 iterations of 10,000 energy measurements with two gradient evaluations. The median final corrected value is $-1.17$.}
    \label{fig:vqe}
\end{figure*}

We proceed as before, with two experiments.
The heatmap generated as the result of the first experiment can be seen in Figure~\ref{fig:heatmaps}.
Through visual inspection of the heatmaps, we see that the compressed case heatmap matches the simulated heatmap more closely and gives us hope of the ground state being found using VQE.
The second experiment confirms that VQE is successful in this encoding, as we see in Figure~\ref{fig:vqe}.
The median final energy of the three corrected VQE runs carried out is $-1.06$.
On the other hand, median final energy of the three uncorrected VQE runs is $-0.90$.
We can both see visually in Figure~\ref{fig:vqe} and in these end values that correction indeed does improve the minimum energy found by the algorithm, bringing it closer to the theoretical minimum.

We next ran VQE using the compressed circuit on a two qubit Aspen-4-2Q-C (the same circuit was used as for Aspen-7-4Q-D; however the CZ gates have higher fidelity on this processor).
The results can be found in Figure~\ref{fig:vqe}.
Notice that correction has a significant impact on the minimal energy found: in the original SPSA algorithm, the median of the final raw (uncorrected) energy value is $-0.96$, however after correction it is  $-1.17$, which is within $6\%$ of the true minimum, $-1.23607$.
The median of corrected final energy value of the three-stage SPSA is found to be $-1.17$ which matches that of the original SPSA algorithm. However the overall number of measurements is smaller, with the traditional SPSA requiring $7 \times 10^6$ circuit evaluations, while the modified needs less than half of that, $2.3 \times 10^6$ circuit evaluations.

\subsection{The measurement of double occupancy}
\begin{figure*}[t]
    \begin{minipage}[t]{0.49\linewidth}
        \includegraphics[width=1.0\linewidth]{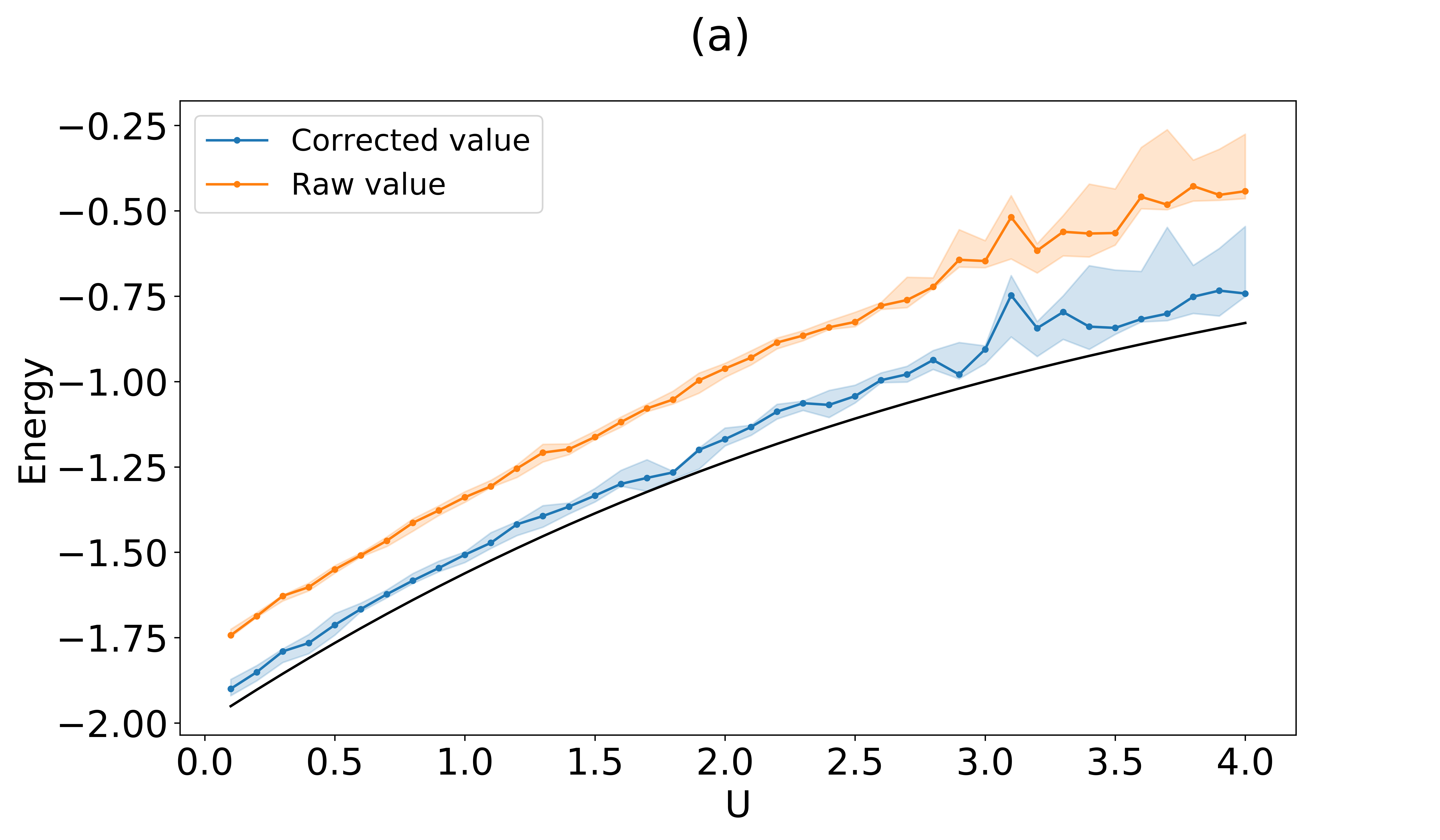}
    \end{minipage}
    \begin{minipage}[t]{0.49\linewidth}
        \includegraphics[width=1.0\linewidth]{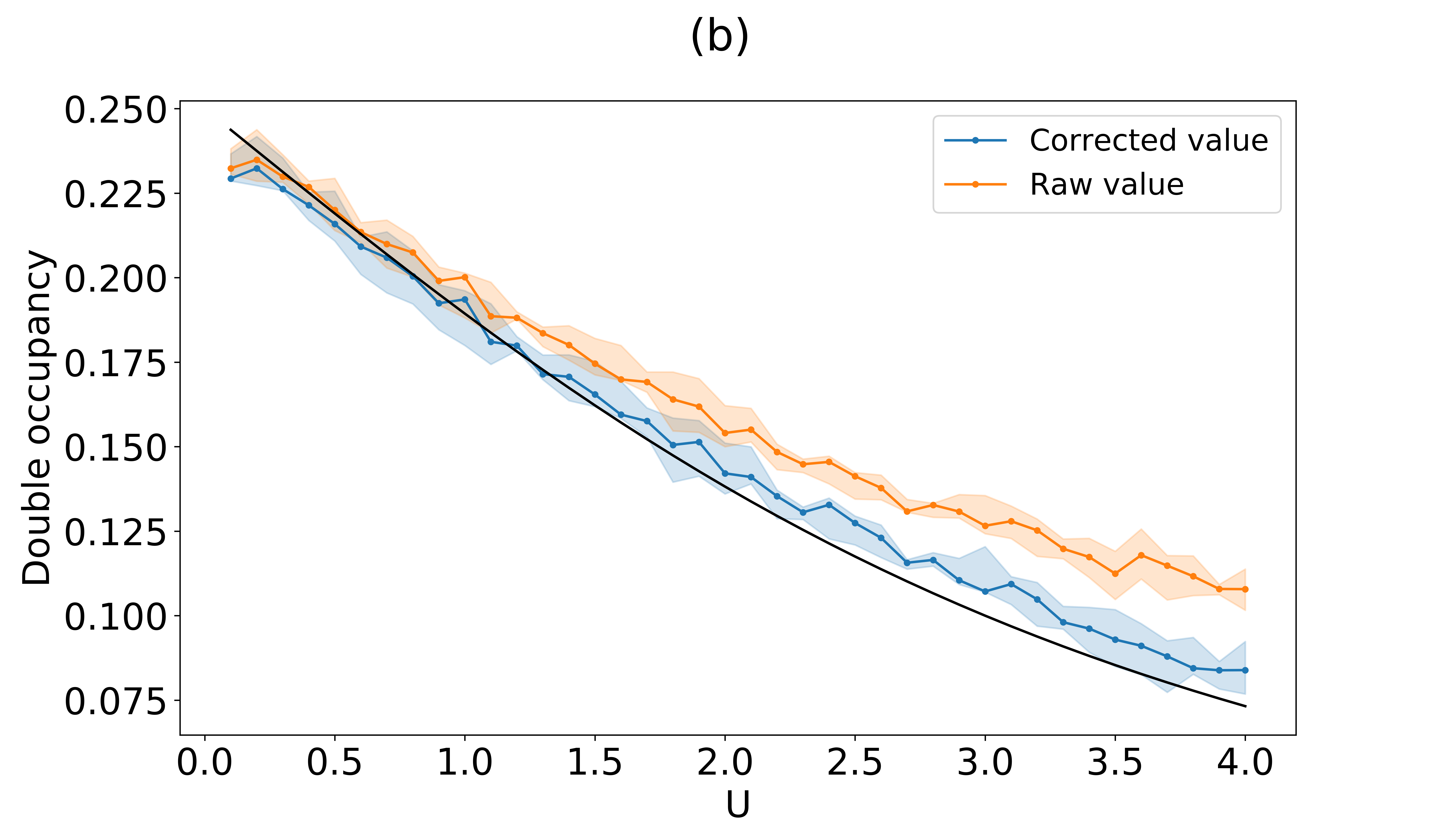}
    \end{minipage}
    \caption{The minimum energy and probability of double occupancy found for a set of different $U$ values. The black line shows the theoretical curves. The blue and orange data points show the median energy value or probability found over five runs, with the envelope giving the minimum and maximum value for each $U$. The orange data points show values with no readout error correction applied, while blue points show values with correction. For each data point VQE was executed with the given $U$ to find the optimal set of parameters, and the more precise last measurement (described in text) was used for the final calculation of both energy and probability values. a) The minimum energy found for a set of different $U$ values. b) The probability of double occupancy for varying $U$ values. }
    \label{fig:variousU}
\end{figure*}

Finally, we extract physically meaningful results from these experiments.
For values of $U \in [0.1, 4]$ in $0.1$ increments, VQE is run five times to find the ground state, and then the energy and probability of double occupancy are extracted.
According to the ground state given in Eq.~\ref{eq:groundstate}, we can calculate this probability to be  $8/\mathcal{N}^2$.
In Figure~\ref{fig:variousU}, we see this theoretical probability (energy) plotted and the median probability (energy) for each $U$ from the processor data.
The envelope shows the minimum and maximum value found in the five runs for each $U$.
The median (25th, 75th percentile) absolute error over $U$ of the median energy values is $6.5 (5.4, 8.7) \times 10^{-2}$ with a maximum error of $2.3 \times 10^{-1}$.
The median (25th, 75th percentile) absolute error over $U$ of the median double occupancy values is $5.7 (3.2, 8.3) \times 10^{-3}$ with a maximum error of $1.4 \times 10^{-2}$.
This is substantially better than the 2-qubit gate fidelity of the processor, which is approximately $96\%$ (and the pyquil estimate of the program fidelity of $83\%$).

Notice the spreading occurring in the envelope as $U$ increases.
This is likely due to a mixture of effects: the onsite component of energy measurement contains a multiplier $U$, which means the error in the onsite measurement will increase as $U$ increases; also, the fidelity of the processor gates changes over time and while the measurements were executed as closely together as possible, they still take a minimum of $8$ hours for the chosen settings. The data presented here were taken over a few days due to other restrictions such as interruptions of access to the processor.

\section{General complexity considerations}
\label{sec:general}

We next argue that the approach we have described for solving compressed instances of the Hubbard model could plausibly demonstrate a quantum speedup over classical computers, when scaled up to larger sizes. To do this, we need to implement the steps required for VQE in time $\poly \log(n)$ for a system of $n$ sites.

First, we show that it is possible to implement measurement and time-evolution according to the terms in $H$ efficiently. The onsite term effectively performs an equality test, which can be implemented using $O(\log n)$ Toffoli and CNOT gates, in depth only $O(\log \log n)$ (assuming arbitrary connectivity). The hopping terms are more complicated. It is natural to split $H_{hop}$ up into blocks of terms that pairwise commute.
These terms can be found by decomposing $G$ into matchings $M \subseteq E$, where a matching is a set of edges that do not share any vertices. Then the Hamiltonian $H_M$ is a direct sum of $X$ matrices acting on 2d subspaces spanned by $\{\ket{i},\ket{j}:(i,j) \in M\}$, so $\braket{\psi|H_M|\psi}$ can be measured by performing Hadamard operations on these subspaces and measuring in the computational basis.

These Hadamard operations in turn can be carried out as follows. For each pair $i<j$ such that $(i,j) \in M$, we want to perform the map $\ket{i} \mapsto \frac{1}{\sqrt{2}}(\ket{i} + \ket{j})$, $\ket{j} \mapsto \frac{1}{\sqrt{2}}(\ket{i} - \ket{j})$. We can do this by first computing the corresponding entry in the matching, so we have $\ket{i}\ket{j}$ (resp.\ $\ket{j}\ket{i}$.) We add an ancilla qubit in the state $\frac{1}{\sqrt{2}}(\ket{0}+(-1)^{[j < i]} \ket{1})$ and, conditional on that qubit being in the state 1, swap the first two registers. We then apply a Hadamard gate again on the last qubit and measure it. With probability $1/2$, the outcome is 0, and the residual state is $\frac{1}{\sqrt{2}}(\ket{i}\ket{j} + (-1)^{[j<i]} \ket{j}\ket{i})$ as desired.

If $G$ is not known in advance, a procedure of Berry et al.~\cite{berry14a} can be used, which decomposes an arbitrary sparse Hamiltonian into terms which have at most one nonzero entry in each row. The number of terms is at most polynomial in the sparsity (which is upper-bounded by a constant for lattice graphs), and a query to a term can be simulated by making $O(1)$ queries to the original Hamiltonian.  

To implement time-evolution according to $H$ (or a subset of its terms), we can use a quantum algorithm for efficient simulation of sparse Hamiltonians. The most efficient such algorithms~\cite{berry15,berry15a,low17} have complexity that scales linearly with the sparsity and evolution time, and logarithmically with the dimension and inverse error.

We stress that it is not necessary to diagonalise $H$ classically (in time $\poly(n)$) to implement these operations efficiently; all that is required is the ability to determine nonzero entries of the interaction graph efficiently (in time $\poly\log(n)$).

\subsection{Classical hardness}

The subspace we consider is of size $O(n)$ for a lattice with $n$ sites, so a classical algorithm could diagonalise the Hamiltonian $H^C$ (or indeed simulate the quantum algorithm directly) in time $\poly(n)$. By contrast, the quantum VQE algorithm uses $O(\log n)$ qubits and could be able to find the ground state in time $\poly\log(n)$. We argue that it is unlikely that this performance could be matched by a classical algorithm on an arbitrary interaction graph.

It was shown by Childs, Gosset and Webb~\cite[Appendix A]{childs14} that approximately computing the lowest eigenvalue of an arbitrary sparse symmetric 0-1 matrix is QMA-complete, implying that it is unlikely that any classical (or indeed quantum) algorithm can do so efficiently in all cases. To be more precise, we assume that we are given access to a sparse $M\times M$ 0-1 matrix $A$ via an efficient classical circuit that takes as input a row (specified by a string of bits), and returns the indices of the nonzero entries in that row. Our goal is to approximate the lowest eigenvalue of $A$ up to accuracy $O(1/\poly\log(M))$.

Finding the ground state energy of the compressed Hubbard model is very similar to this problem. Consider the variant of the problem where we generalise in two ways: firstly, the interaction graph on $n$ vertices is unknown in advance, can be an arbitrary sparse graph, and is determined via an efficient classical subroutine; secondly, the Hamiltonian $H$ is generalised to $H_w$ by including an arbitrary weighted one-body term of the form $H_{loc} = \sum_i w_i n_{i\uparrow}$ (and similarly for spin-down). It is easy to see that the compressed Hamiltonian $H^C_{loc}$ is precisely an arbitrary diagonal matrix. In the special case $U=0$, the lowest eigenvalue of $H_w^C$ corresponds to the lowest eigenvalue of $H^C_{hop} + H^C_{loc}$, which is an arbitrary sparse symmetric matrix.

This puts us in the setting of Childs, Gosset and Webb's work, exponentially compressed by replacing $M$ with $n$; put another way, the problem of approximating the ground state energy of $H_w^C$ is $\operatorname{QMA}$-complete for a system of logarithmically many qubits.
This suggests that for arbitrary sparse interaction graphs, no classical -- or indeed quantum -- algorithm could find the ground state energy in time $\poly \log(n)$. However, the hope, as with variational quantum algorithms in general, is that for some Hamiltonians, a quantum algorithm could find the ground state energy where a classical algorithm cannot.

Finally, we remark that if we generalise further to allow arbitrary weights on the hopping terms $a_i^\dag a_j$ too, the compressed Hamiltonian $H_{hop}^C$ can be an arbitrary Hermitian $n\times n$ matrix, implying that it can simulate an \emph{arbitrary} spin Hamiltonian on $\log_2 n$ qubits.

\subsection*{Acknowledgements}

We would like to thank Alan Aspuru-Guzik, Ryan Babbush, Zhang Jiang, Lana Mineh and Sabine Tornow for helpful comments on the paper, Toby Cubitt for discussions about compressed representations of Hamiltonians, and Rigetti for access to their Aspen-4 and Aspen-7 systems.

\bibliographystyle{mybibstyle}
\bibliography{main,strategies}

\end{document}